\documentstyle[aps,prl,multicol,epsfig]{revtex}
\begin{document}
\title{ Thermally generated vortices, gauge invariance and
      electron spectral function in the pseudo-gap regime}
\author{  Jinwu Ye  }
\address{
   Department of Physics, The Pennsylvania State University, University Park, PA, 16802 }
\date{\today}
\maketitle
\begin{abstract}
    Starting from classical vortex fluctuation picture,
    we study the single electron
    properties in the pseudogap regime.
    We show that it is the gauge invariant Green function of spinon which is
    directly related to ARPES data in the pseudogap regime
    instead of the non-gauge invariant one.
    We find that the random gauge field from the
    thermally generated vortices completely
    destroys the coherent spinon motion and leads to
   collective excitations pertinent to non-Fermi liquid behaviors.
   The Energy Distribution Curves (EDC) show broad peaks, while
   the Momentum Distribution Curve (MDC) show rather sharp peaks with
   Lorenz form. The local density of state at zero energy 
   scales as the inverse of
   Kosterlize-Thouless length. These results are qualitatively
    consistent with the ARPES data in the pseudo-gap regime.
\end{abstract}
\begin{multicols}{2}

    Although the superconducting state of high $ T_{c} $ cuprates
    can be described approximately as  pair condensates with a d-wave
    symmetry and well-defined quasi-particle excitations above them,
    the anomalous properties of the normal state 
    are still poorly understood.
   Many experiments such as Angle-Resolved Photoemission (ARPES)
   \cite{ding}, Scanning Tunneling Microscopy (STM) \cite{ren}, Inelastic Neutron
   Scattering experiments \cite{dai} and
   transport characteristics \cite{corson,ong} provided evidence for the
   opening of a gap in the electronic excitation spectrum 
   above the critical temperature $ T_{c} $ up to
   some characteristic temperature $ T^{*} $ in the underdoped
   cuprates, this energy gap is dubbed " pseudo-gap" ( Fig.1).
   Both ARPES \cite{ding} and STM \cite{ren}
   have indicated that in the pseudo-gap regime, the
   pseudogap possesses the same $ d_{x^{2}-y^{2}} $ 
   symmetry as the superconducting gap
   below $ T_{c} $. ARPES detected sharp quasi-particle 
   peak below $ T_{c} $, but only broad spectral weight above $ T_{c} $.

\vspace{0.2cm}

\epsfig{file=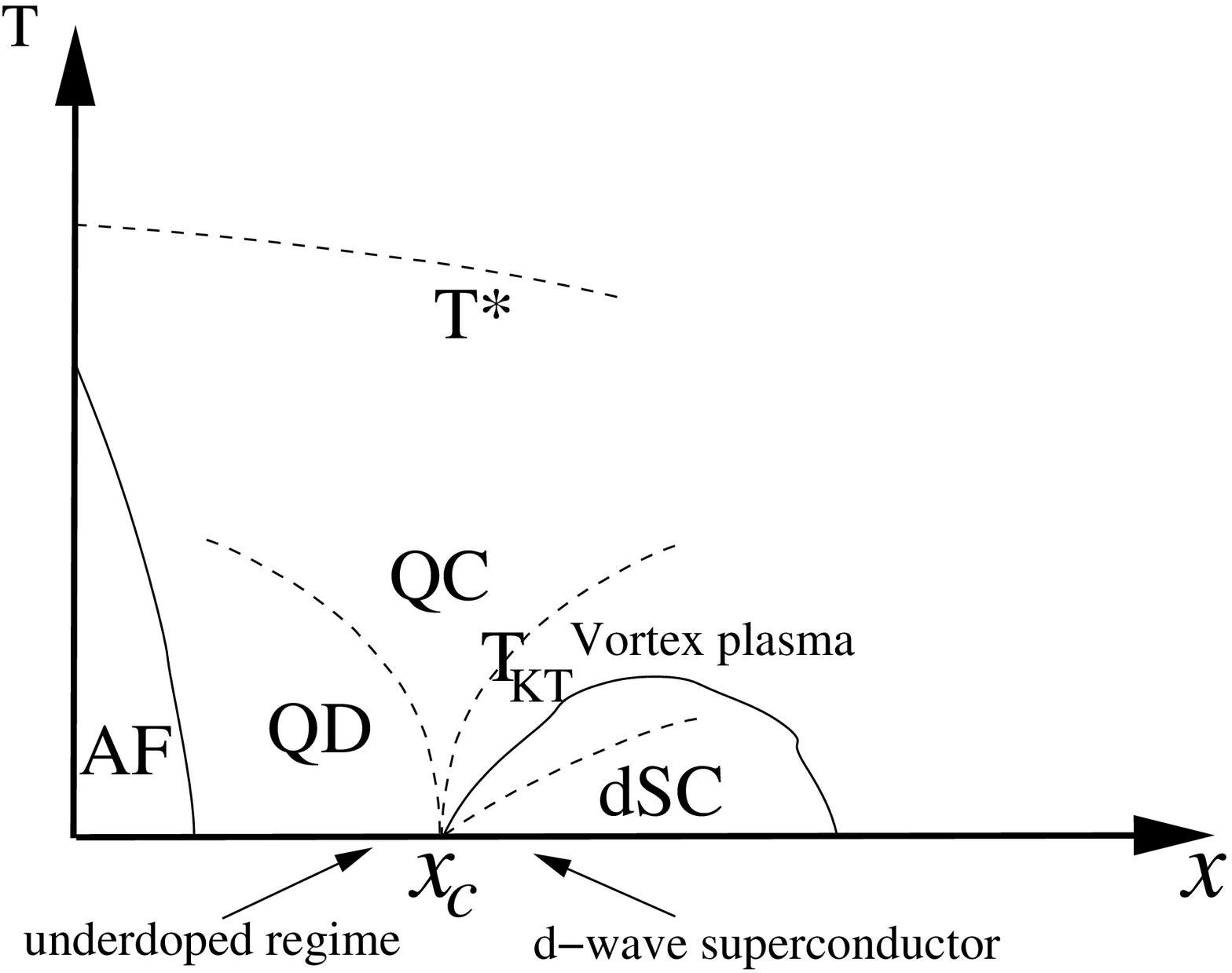,width=3.2in,height=2.2in,angle=0}

{\footnotesize {\bf Fig 1:} There is a quantum phase transition from
  d-wave superconductor to the underdoped regime at $ x=x_{c} $.
  The Quantum Critical (QC), Quantum Disordered (QD) and 
  Vortex plasma regimes are defined
  in \cite{subir}. There is {\em no} spin-charge separation in QD and QC
  regimes. But there is approximate spin-charge separation in the vortex
  plasma regime (see the text). }

\vspace{0.25cm}

   In this paper, we are trying to explain the ARPES data 
   near the four nodes $ (\pm 0.39 \pi, \pm 0.39 \pi ) $ \cite{not} from the
   physical picture proposed by Emery and Kivelson and others
   \cite{emery,carl,fisher}:
   the finite temperature superconducting transition is a
   Kosterlize-Thouless (KT) transition driven by vortex pair un-binding,
   namely $ T_{c}= T_{KT} $. Below $ T_{KT} $, the phase fluctuation is small,
   the positive and negative vortex pairs are tightly bound
   together, above $ T_{KT} $, the long-range superconducting
   coherence is destroyed by strong phase fluctuation which leads to
   the generation of free vortices with both parity, but
   the amplitude of the {\em local} superconducting order parameter remains
   non-vanishing up to $ T^{*} $.  The quasi-particles in the pseudo-gap regime
   are the remnants of those in the superconducting phase and are
   strongly scattered by the free vortices above $ T_{KT} $.
   We are trying to calculate 
   the {\em real electron} spectral weight due to these scatterings and
   compare with the experiments.

    The quasi-particle scattering due to the Volovik effect \cite{vol}
  in the pseudogap 
  regime was discussed in Refs.\cite{andy}. As first stressed in Ref.
  \cite{and}, there is also Aharonov and Bohm (AB) phase effect from
  $ hc/2e $ vortices. Although the AB phase effect only provides
  a periodic potential
  in the vortex lattice states discussed in \cite{and,bert},
  it starts to scatter quasi-particles when the vortices are randomly 
  distributed.
  As pointed out in \cite{jinwu}, the elastic scattering due to the AB phase
  from the disordered vortex array dominate over that from the
  Volovik effect. In \cite{tough}, Millis and the author studied 
  the zero temperature quantum phase transition from
  d-wave superconductor to the underdoped regime driven by the condensation
  of quantum generated $ hc/2e $ vortices. In this paper, we apply the method developed in
  Refs.\cite{jinwu,tough} to study the electron spectral weight
  in the classical vortex fluctuation regime ( the vortex plasma regime in
  Fig.1 ). We derive the important Eq.\ref{qed}
  by two approaches. First, replacing vortex density $ n_{v} $ 
  generated by external magnetic field in \cite{jinwu} by the free 
  vortex density $ n_{f}(T) $ above $ T > T_{KT} $, we lead to Eq.\ref{qed}
  directly from the results in Ref.\cite{jinwu}. Second, starting from
  the zero temperature effective action developed in Ref.\cite{tough},
  we show that the action reduces to Eq.\ref{qed} after treating
  the vortices classically.  We show that the random gauge field $ a_{\alpha} $
  in Eq.\ref{qed} from the thermally generated vortices
  completely destroys the coherent quasi-particle picture and leads to
  collective excitations pertinent to Non-Fermi liquid character.
  The Energy Distribution Curves (EDC) ( at fixed $ k $ )
  show broad peaks, while
  the Momentum Distribution Curve (MDC) ( at fixed $ \omega $ )
  show rather sharp peaks with Lorenz form.
  Volovik effect leads to finite local DOS scaled as the inverse of the KT correlation length.
  These results are qualitatively
  consistent with the ARPES data in the pseudo-gap regime.
  The ARPES spectrum in QC and QD regimes will be discussed elsewhere \cite{un}.

  In Ref.\cite{jinwu}, starting from BCS Hamiltonian, performing the singular gauge transformation
  and expanding $ H_{s} $ around the node 1 where
     $ \vec{p}=(p_{F},0) $, Ye obtained $ H_{s}
     =H_{l}+H_{c} $ where the linearized Hamiltonian
     $ H_{l} $ is given by \cite{linear}:
\begin{eqnarray}
   H_{l} & = &  v_{f} (p_{x}+  a^{\psi}_{x} ) \tau^{3} 
   +v_{\Delta} (p_{y}+  a^{\psi}_{y} ) \tau^{1} 
   + v_{f}  v_{x}(\vec{r})   \nonumber  \\
  & + &  (1 \rightarrow 2, x \rightarrow y )
\label{linear}
\end{eqnarray}
   where $ \vec{v}_{s}= \frac{1}{2} (\vec{v}^{A}_{s}+ \vec{v}^{B}_{s})
   = \frac{\hbar}{2} \nabla \phi-\frac{e}{c} \vec{A}$ is the total
     superfluid {\em momentum} and
 $ a^{\psi}_{\alpha}= \frac{1}{2}( v^{A}_{\alpha}-v^{B}_{\alpha})
 = \frac{1}{2}( \nabla \phi_{A}-\nabla \phi_{B}) $ is the
   internal gauge field.
    We get the corresponding expression at node $ \bar{1} $ and $ \bar{2} $
  by changing $ v_{f} \rightarrow -v_{f}, v_{\Delta} \rightarrow -v_{\Delta} $
  in the above Eq. For explicit spin $ SU(2) $ invariant formulation,
  see Ref.\cite{tough}.

    The curvature term $ H_{c} $ can be written as:
\begin{equation}
   H_{c} =  \frac{1}{m} [  \{ \Pi_{\alpha}, v_{\alpha} \}
	  +  \frac{  \vec{\Pi}^{2}+ \vec{v}^{2} }{2} \tau^{3} 
	  + \frac{\Delta_{0}}{2 \epsilon_{F}} \{ \Pi_{x}, \Pi_{y}\} \tau^{1} ]
\label{curv}
\end{equation}
     Where $ \vec{\Pi}= \vec{p}+  \vec{a}^{\psi} $ is the covariant
     derivative. $ H_{c} $ takes the {\em same } form for all the four nodes \cite{quit}.

 Just like at filling factors at $ \nu=1/3 $ or $ \nu=1/2 $ QH system\cite{hlr},
 the transport properties 
 can be studied directly in the transformed
 Hamiltonian. But the single particle {\em electron}
 Green function is much more difficult to calculate \cite{he}.
 In the present problem, the general form of electron annihilation operator
 was given in Ref. \cite{tough}:
\begin{eqnarray}
   C_{\alpha}(\vec{x}) & = & \sum_{i=1,2} [ e^{i \phi/2 }
    e^{i \vec{K}_{i} \cdot \vec{x}} e^{i \int a^{\psi} d x} \psi_{i 1 \alpha}
                  \nonumber  \\
  & - & \epsilon_{\alpha \beta}
    e^{i \phi/2 } e^{-i \vec{K}_{i} \cdot \vec{x}}
    e^{-i \int a^{\psi} d x} \psi^{\dagger}_{i 2 \beta} ]
\label{single}
\end{eqnarray}
     where $ i $ is the node index, $ 1, 2 $ are $p-h $ indices and
    $ \alpha, \beta $ are spin indices.

 All the above equations can be applied straightforwardly
 to the pseudogap regime above $ T > T_{KT} $ where vortices are unbound.
 There are equal number of {\em free}  positive and negative 
 vortices generated by
 thermal fluctuation. On the average, the vanishing of $ v_{\mu} $ and
 $ a^{\psi}_{\mu} $ are automatically ensured.
 Furthermore, the thermal generated vortices are moving around and their 
 smearing effect makes the $ U(1) $ character stronger than their $ Z_{2} $
 character in randomly-pinned static vortex array case. 
 Replacing vortex density $ n_{v} $ generated by external
 magnetic field in \cite{jinwu} by the free vortex density $ n_{f}(T) $ above
 $ T > T_{KT} $, we can describe the effective action  
 by spinons moving in a {\em static} magnetic field
 ( there is no feedback effects on the gauge field propagator from the fermions )
   \cite{jinwu,tough}:
\begin{equation}
   {\cal L} =  \psi^{\dagger}_{a} \gamma_{\mu} (\partial_{\mu}
    -i  a^{\psi}_{\alpha} ) \psi_{a} + \frac{1}{4 n_{f} } 
   ( f^{\psi}_{\alpha \beta} )^{2}
\label{qed}
\end{equation}
  Where $ \alpha, \beta=1,2 $ are the space components only;
  $ a=1,\cdots, N=4 $ are 4 species of Dirac fermion and $
  n_{f}(T) \sim \xi^{-2}(T) $ where $ \xi(T) $ is the KT correlation
  length at $ T > T_{c} $.

  In Ref.\cite{tough}, the authors derived an effective action describing
  the zero temperature quantum phase transition from d-wave superconductor
  to underdoped regime around the critical doping $ x_{c} $.
  At zero temperature, the vortices have to be treated
  {\em quantum } mechanically, one has to perform {\em two} singular
  transformations which 
  are dual to each other to quasi-particles and moving vortices 
  respectively to keep all the possible commutation relations intact. 
  Just like conventional singular gauge transformation \cite{hlr} leads to
  conventional Chern-Simon (CS) term, the two mutual singular gauge 
  transformations lead to a mutual CS term. In the effective action,
  quantum generated vortices couple to quasi-particles by the mutual CS term,
  the vortices are also interacting with a charge $ U(1) $ gauge field $ a_{\mu} $
  mediating the long-range {\em current-current} logarithmic interactions between the vortices.
  The $ U(1) $ charge fluctuation leads to mass terms not only
  for $ v $, but also for the spatial component $ \vec{a}^{\psi} $.
  There is {\em no gapless dynamic } gauge fluctuations in the Cooper-pair
  picture. The precise nature of the zero temperature quantum critical point (QCP)
  between the d-wave superconductor and some unknown charge ordered state
  is still under investigation.
  The electron spectral functions in the QC and QD regimes are controlled by
  this QCP  and will be presented elsewhere \cite{un}.

  However, the spectral function in the vortex plasma regime around $ T_{KT} $ in Fig.1
  are determined by the classical hydro-dynamics of vortices which
  is the focus of the present paper.  The vortices being treated classically \cite{cmr}, their
  commutation relations can be neglected, the dual singular gauge 
  transformation in the vortices is not necessary, the Berry phase term for
  the boson ( the linear derivative term) can also be neglected \cite{cmr}.
  More specifically, only the time component of the {\em charge} gauge field $ a_{0} $ 
  is kept to mediate the long-range {\em density-density}
  logarithmic interaction between the vortices. Obviously, $ a_{0} $ couples
  to the vortices the same way as the superfluid velocity $ v_{\alpha} $
  couples to the spinon,
  its fluctuation leads to a mass term for the $ v-v $ correlation.
  However, being orthogonal to the spatial component
  $ \vec{a}^{\psi} $, $ a^{\psi}-a^{\psi} $
  correlator remains gapless and is given by the Maxwell term in Eq.\ref{qed}.
  This is exactly consistent with the results
  first derived in \cite{jinwu}.

  It is important to point out that what measured by ARPES is the spectral
  function of the real electron in Eq.\ref{single} which carry both spin and
  charge, instead of that of the spinon $ \psi $ which carry spin only.
  It is vital to emphasize that the electron Green function 
  $ G(\vec{x}, t) =< C_{\alpha}(0,0) C^{\dagger}_{\alpha}(\vec{x}, t) > $ is 
  gauge invariant under the internal gauge transformation \cite{tough}, 
  but that of the spinon $ \psi $ is {\em not} !
  We make the following {\em gauge invariant } mean field approximation for
  the electron Green function:
\begin{eqnarray}
   & & G(\vec{x}, t)  =  < e^{i \phi(\vec{x})/2} e^{-i \phi(0)/2 } >
                                      \nonumber  \\
  & [ &  e^{i \vec{K}_{i} \cdot \vec{x}}
    < \psi_{i 1 \alpha}(\vec{x},t) e^{i \int^{x}_{0} a^{\psi}(x) dx}
    \psi^{\dagger}_{i 1 \alpha}(0,0)>   \nonumber  \\
   & + & e^{-i \vec{K}_{i} \cdot \vec{x}}
   < \psi^{\dagger}_{i 2 \alpha}(\vec{x},t) e^{ -i \int^{x}_{0} a^{\psi}(x) dx}
   \psi_{i 2 \alpha}(0,0)> ]
\label{e}
\end{eqnarray}
 
    As shown in Ref.\cite{jinwu}, it is vital to make a mean field
  approximation to respect the exact T symmetry. Very similarly, 
  in the pseudo-gap
  regime, it is equally important to make a mean field approximation to respect
  the exact  gauge invariance.
  {\em Any approximation violating the gauge invariance will lead to
  completely wrong conclusions !}
    The vortices being treated classically,
   only the $ \omega =0 $ component is kept. So the vortices fluctuate
   much slowly than the fermions which are treated completely quantum 
   mechanically. We expect the 
  the spin-charge separation indicated in Eq.\ref{e} is a good approximation
   in the vortex hydro-dynamic regime.
   However, as shown in \cite{tough},
   there is no spin-charge separation at $ T=0 $, in QC and QD regimes (Fig.1).
   This is in contrast to $ Z_{2} $ gauge
  theory  \cite{z2} where there is {\em true} spin-charge separation
   at $ T=0 $.

  The first factor in Eq.\ref{e} is the classical correlation function of the vortices,
  at $ T > T_{KT} $, it decays exponentially:
\begin{equation}
    G_{v}(\vec{x}) = < e^{i \phi(\vec{x})/2} e^{-i \phi(0)/2 } >
    \sim e^{-x/ \xi}
\end{equation}

    The second factor in Eq.\ref{e} is the {\em gauge invariant} Schwinger
  Green function of the spinon in a given configuration of a static gauge
  field:
\begin{equation}
  [ G^{inv}_{s}(\vec{x},t)]_{11}
  = < \psi_{i 1 \alpha}(\vec{x},t) e^{i \int^{x}_{0} a^{\psi}(x) dx}
    \psi^{\dagger}_{i 1 \alpha}(0,0)>
\label{inv}
\end{equation}

  The physical meaning of Schwinger gauge invariant Green function
  has been painfully sought. Remarkably, it is this function directly
  involved in the ARPES in the pseudogap regime in
  high temperature superconductor  \cite{wen} !

  It is well known that the infrared divergence in the gauge fluctuation in
  Eq.\ref{qed} plagues the non-gauge invariant spinon Green function 
  $ G_{s}(\vec{x},t) = 
  < \psi_{i 1 \alpha}(\vec{x},t) \psi^{\dagger}_{i 1 \alpha}(0,0)> $. As shown in \cite{jinwu},
  it leads to divergent quasi-particle scattering rate  $ 1/ \tau_{l} $.
  However, the infra-red divergence has {\em no } physical meaning and
  was shown to be cancelled by the vertex correction in the gauge invariant transport time 
  $ \tau_{tr} $ \cite{jinwu}. Very similarly, this divergence will 
  disappear in the gauge
  invariant $ G^{inv}_{s}(\vec{x},t) $ which directly determine the ARPES data.
   The motion of non-relativistic particle in random magnetic field
   has been studied in Ref.\cite{static}. It was found that the particle's
  propagation in random magnetic field ceases to be coherent and 
  $ G^{inv}_{s}(\vec{x},t) $ develops {\em branch cut singularities}.
  Therefore quasi-particle picture completely breaks down which is a hallmark
  of non-Fermi liquid picture. Studying {\em massless} Dirac fermion
  in random magnetic problem is a much more complicated problem, we leave
  it to future publication \cite{un}.
  Instead, we can look at a much simpler
  problem of Dirac fermion interacting with  {\em dynamic} fluctuating gauge 
  field. This simpler problem has Lorenz invariance at $ T=0 $ which
  dictates $ G^{inv}_{s}(k) \sim i k_{\mu} \gamma_{\mu}
  /k^{2-2\eta} $ with $ k=( \vec{p}, i \omega_{n} ) $.
   It can be explicitly written as
\begin{equation} 
   G^{inv}_{s}(\vec{p},i \omega_{n})
      =\frac{1}{ [ E^{2}_{\vec{p}}-( i\omega_{n})^{2} ]^{1-\eta} }
     \left ( \begin{array}{cc}
		i \omega_{n} + v_{f} p_{x}   &   v_{\Delta} p_{y}   \\
		v_{\Delta} p_{y}   &   i \omega_{n} - v_{f} p_{x}  \\
		\end{array}   \right )    
\label{cut}
\end{equation}
    where $ E^{2}_{\vec{p}} =( v_{f} p_{x} )^{2} + ( v_{\Delta} p_{y} )^{2} $
   and $ \eta $ is the anomalous dimension which was calculated to one loop
    in $ 1/N $ expansion to be $ \eta \sim 0.27 $ in Ref.\cite{wen}.

  Taking analytic continuation to the real frequency and then taking imaginary
  part, we get the spectral function $ A( \vec{k}, \omega )= - Im G ( \vec{k}, \omega + i \epsilon ) $
  of the electron near $ \vec{K}_{i} $ at temperature $ T $ :
\begin{eqnarray}
  A(\vec{k}, \omega)= \int \frac{d^{2} \vec{p}}{ (2 \pi)^{2} }
    \frac{ \sin (\pi \eta) }{ (\vec{k}-\vec{p} )^{2}+ \xi^{-2}(T) }
    \frac{\omega + v_{f} p_{x} }
    { [ \omega^{2}- E^{2}_{\vec{p}} ]^{1-\eta} }  \nonumber   \\
    \times
    [ \theta( \omega-E_{\vec{p}} ) - \theta(- \omega-E_{\vec{p}} ) ] ~~~~
\label{spectral}
\end{eqnarray}
    with the similar expression at $ -\vec{K}_{i} $.

  It can be shown that
\begin{eqnarray}
  A(\vec{k}, \omega \rightarrow 0)
  & = & \frac{sin (\pi \eta) | \omega |^{1+2 \eta}}
                       {4 \pi (k^{2}+ \xi^{-2}(T))} 
                              \nonumber  \\ 
   A(\vec{k}, \omega \rightarrow \infty )
   & = &  \frac{sin (\pi \eta) |\omega|^{-1+2 \eta} }{4 \pi \eta }
\end{eqnarray}   
  Precise energy distribution curve (EDC) ( at fixed $ k $)
  and momentum distribution curve (MDC) ( at fixed $ \omega $ ) 
   for $ \xi(T) \sim 10 $ are obtained numerically and drawn in Fig.2.

\vspace{0.25cm}

\epsfig{file=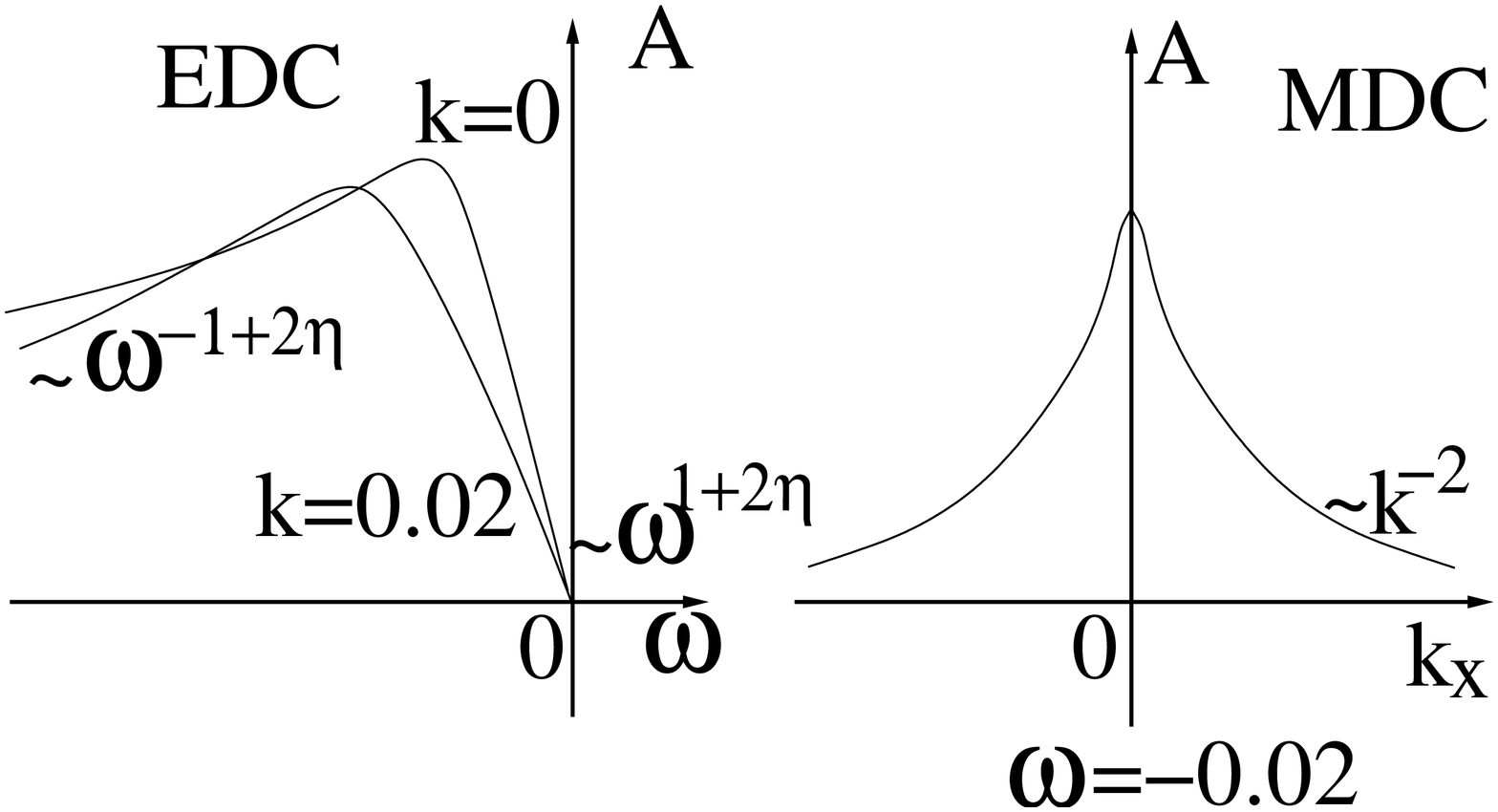,width=3.2in,height=1.8in,angle=0}

{\footnotesize {\bf Fig 2:} Energy distribution curve (EDC)
  ( at fixed $ k $)  and Momentum distribution curve (MDC)
  ( at fixed $ \omega $ ). }

\vspace{0.25cm}
 
  It is easy to see that just as random static gauge field, the dynamically
  fluctuating gauge potential also destroys the coherent quasi-particle picture.
  The spinon in the pseudo-gap regime loses its identity and replaced by collective excitations,
  in contrast to its counter-part in the superconducting state. 
   The Volovik effect is short-ranged as shown in \cite{jinwu},
   its effect is far less dramatic than that of the random gauge field
   which completely destroys the quasi-particle picture.
   Our interpretation of the ARPES data is in sharp contrast with
   that of $ Z_{2} $ gauge theory \cite{z2}. In this theory, the gauge field
   is absent due to the generation of {\em double} strength $ hc/e $
   vortices which leads to spin charge separation in the
    long length scale at $ T=0 $.

    From the long time behavior of $ G(\tau) \sim G^{inv}_{s} (\tau )
    \sim 1/ \tau^{2+2\eta} $, we find the local density of state (DOS) 
    $ N(\omega) \sim \omega^{1+ 2 \eta } $. However, as shown in \cite{jinwu}
    the Volovik effect will generate finite DOS at zero energy
    $ N( \omega=0, T) \sim \sqrt{ n_{f} } \sim \xi^{-1}(T) $.
    So the EDC curves in Fig. 2 should
    approach a small finite value at very small frequency
    due to the Volovik effect.
  
  A few caveats should be pointed out before comparing Fig.1 with 
  the available ARPES data in the node directions:(1) The mean field
  approximate spin-charge separation Eq.\ref{e} is only valid at vortex
  hydro-dynamic regime ( Fig.1 ). It is only $ 30-40 K $ above $ T_{KT} $ consistent with
  the recent experiments by Corson {\sl et al}. Raising temperature higher will move into
  QC regime. There is no spin-charge separation at $ T=0 $. A different calculation is needed
  to calculate ARPES spectrum in QC and QD regimes \cite{un} 
  (2) Conceptually, the gauge invariant Green function should be calculated in a static gauge field 
  instead of in a dynamic gauge field. However, we do not expect the technical calculation
  in static gauge field will lead to dramatically different EDC and MDC than those
  in Fig.2, because unlike the gauge non-invariant Green function, the gauge invariant
  one is free of infra-red divergences. Therefore, we expect Fig.2 
  qualitatively describe the electron spectral weight in the vortex hydro-dynamic regime if
  the phase fluctuation scenario is the correct one.

    I thank A. J. Millis for very helpful discussions throughout the paper.
    I am indebted to J. K. Jain and Y. Liu for discussion on vortex dynamics.
    I am grateful for X. G. Wen to explain their unpublished results to me.
    I also thank B. Halperin, R. Jackiw, M. Ma, T. K. Ng 
    and C. Ordonez for helpful discussions and Y. Chen for the helps on the Fig.2.
    I also thank the two anomynous referees for careful reading and helpful comments
    on the paper. This work was supported by Penn State University.

\end{multicols}

\begin{references}
\bibitem{not} The ARPES data near the maximum gap
     $ (\pi,0) $ and $ (0, \pi) $ will be presented elsewhere. 
     The quasi-particles in these anti-node directions will also
     couple to the $ ( \pi, \pi ) $ resonance mode detected by INS.
\bibitem{linear} The maximum gap is $ \sim 300K $, we expect the
     linearization is a reasonable approximation when $ T \sim 100K $.
\bibitem{quit}  Due to the exact Time Reversal symmetry
  in the pseudo-gap regime in zero
  magnetic field, all the skewnesses defined in \cite{jinwu} vanish,
  therefore $ \kappa_{xy} $ vanishes.
  We will not discuss the curvature term $ H_{c} $ in the following.
\end{references}
\end{document}